\documentclass[conference]{IEEEtran}
\usepackage[utf8]{inputenc}
\usepackage{amsmath}
\usepackage{amsfonts}
\usepackage{amssymb}
\usepackage{graphicx}
\usepackage{cite}
\usepackage{url}
\usepackage{hyperref}
\usepackage{tikz}
\usetikzlibrary{positioning,arrows.meta,shapes.geometric}

\title{Quantum Artificial Intelligence for Secure Autonomous Vehicle Navigation: An Architectural Proposal}

\author{
\IEEEauthorblockN{Hemanth Rajesh K}
\IEEEauthorblockA{University of Rhode Island and QAIgs.com\\
Hemanth.kanna@uri.edu, raj.hemanth@qiags.com}
\and
\IEEEauthorblockN{Sowmya Chintalpaudi}
\IEEEauthorblockA{Carleton University and QAIgs.com\\
sowmyachintalapudi@cmail.carleton.ca}
}

\begin{document}

\maketitle

\begin{abstract}
Navigation is a very crucial aspect of autonomous vehicle ecosystem which heavily relies on collecting and processing large amounts of data in various states and taking a confident and safe decision to define the next vehicle maneuver. In this paper, we propose a novel architecture based on Quantum Artificial Intelligence by enabling quantum and AI at various levels of navigation decision making and communication process in Autonomous vehicles : Quantum Neural Networks for multi-modal sensor fusion, Nav-Q for Quantum reinforcement learning for navigation policy optimization and finally post-quantum cryptographic protocols for secure communication. Quantum neural networks uses quantum amplitude encoding to fuse data from various sensors like LiDAR, radar, camera, GPS and weather etc., This approach gives a unified quantum state representation between heterogeneous sensor modalities. Nav-Q module processes the fused quantum states through variational quantum circuits to learn optimal navigation policies under swift dynamic and complex conditions. Finally, post quantum cryptographic protocols are used to secure communication channels for both within vehicle communication and V2X (Vehicle to Everything) communications and thus secures the autonomous vehicle communication from both classical and quantum security threats. Thus, the proposed framework addresses fundamental challenges in autonomous vehicles navigation by providing quantum performance and future proof security.
\end{abstract}


\section{Introduction}

The global autonomous vehicle industry, valued at \$1.5 trillion in 2022, is expected to be $\sim$\$13.63 trillion by 2030, representing a compound annual growth rate of 32.3\%. This revolutionary shift in transportation depends critically on accurate navigation systems that fuse multi-modal sensor data, including LiDAR, camera, radar, GPS, and weather information. However, current systems encounter substantial challenges in achieving robust and secure sensor fusion, particularly under adversarial attacks that manipulate sensor inputs.

Autonomous vehicles require numerous sensors for safety and efficiency, yet classical multi-modal fusion methods struggle with sensor misalignment, noise \cite{polychronopoulos2007sensor}, and real-time processing constraints, directly impacting navigation safety. Furthermore, classical encryption schemes in autonomous vehicles are becoming vulnerable to emerging quantum computing threats. This vulnerability not only compromises data confidentiality but also threatens civilian safety as autonomous vehicles scale exponentially.

In lieu of potential quantum threats, post-quantum cryptography (PQC) algorithms are becoming popular to secure future systems. PQC algorithms are already securing Vehicle-to-Vehicle (V2V) and Vehicle-to-Everything (V2X) communication links \cite{twardokus2024cryptography}. However, current research does not address vulnerabilities in intra-vehicle sensor-to-processor communication channels.

This work proposes a comprehensive Quantum Artificial Intelligence (QAI) architecture that integrates quantum neural networks for sensor fusion, Nav-Q quantum reinforcement learning for navigation policy optimization, and post-quantum cryptographic protocols for secure communication. By embedding QNN-based sensor fusion with Nav-Q policy learning and PQC-secured data transmission into an integrated pipeline, our framework provides quantum-enhanced navigation that could secure future autonomous vehicle ecosystem against both classical and quantum threats. This establishes the foundation for next-generation autonomous vehicle systems that leverage the full potential of quantum artificial intelligence.

\section{Related Work}

\subsection{Quantum Sensor Fusion in Autonomous Vehicles}

Sensors are the neural network of autonomous vehicle systems. Considering the multitude of sensors in decision-making, sensor fusion represents a major aspect of autonomous vehicle perception systems, traditionally combining radar, LiDAR, camera, GPS, and weather data through classical neural networks. However, real-time synchronization has challenges due to high-dimensional representations and misaligned data streams.

Zhou et al. proposed a quantum-enhanced sensor fusion approach using quantum computational methods for vehicle perception systems, demonstrating potential advantages in classification tasks. Their work provides foundational insights into quantum sensor processing, though focused on classical machine learning integration rather than quantum neural network architectures. Our QNN approach builds on these quantum sensing principles while implementing full quantum neural network processing for comprehensive multi-modal fusion.

Though there are several advances in quantum-enhanced perception, current work still has limitations. First, current systems focus heavily on static perception tasks rather than navigation policy learning, limiting their ability to model sequential decision-making under dynamic conditions. Second, the scope of sensor data integration remains narrow compared to requirements for robust navigation. Finally, existing work does not consider how fused quantum data representations integrate with downstream learning policies or system behavior under adversarial conditions. Our proposed framework addresses these limitations by expanding the sensor dataset, integrating variational circuits into policy learning, and embedding security and adversarial robustness layers.

\subsection{Nav-Q - Quantum Reinforcement Learning for Autonomous Vehicle Navigation}

Deep reinforcement learning (DRL) is a widely adopted technique in autonomous and semi-autonomous vehicle research. It enables continuous decision making based on sensor inputs. As quantum computing evolves, the integration of quantum processing with deep reinforcement learning offers potential for optimizing navigation strategies under complex constraints. Recent work in quantum reinforcement learning has explored the application of variational quantum circuits for navigation and control tasks. This work shows promise for collision-free autonomous vehicle navigation. These approaches demonstrate the viability of quantum policy gradient optimization, though typically operating on simplified sensor inputs without integration with quantum sensor fusion or comprehensive security protocols. Our work extends quantum reinforcement learning by integrating it with QNN-based sensor fusion and post-quantum security, creating a comprehensive quantum AI pipeline where quantum circuits process quantum-fused sensor representations rather than classical inputs.

\subsection{Quantum Adversarial Machine Learning}

Quantum machine learning models show promise in both computational capacity and potential advantages, yet they exhibit vulnerability to adversarial attacks that cause incorrect outputs through subtle input alterations. West et al. \cite{west2023towards} reviewed the emerging field of Quantum Adversarial Machine Learning (QAML) and analyzed how classical attack strategies like Fast Gradient Sign Method (FGSM) and Projected Gradient Descent (PGD) attacks adapt to quantum circuits. They introduced a classification of QAML threat models, highlighted challenges of adversarial robustness in hybrid quantum-classical models, and proposed initial defense strategies including randomized encoding and input purification.

However, their study remains largely theoretical, focusing on quantum classifiers in abstract tasks like image classification without extension to real-world autonomous vehicle systems. Specifically, their experiments and threat models operate in simulated environments, limiting applicability to real-world attack modalities like LiDAR spoofing or GPS jamming, which represent primary attack surfaces in autonomous vehicles. Additionally, the work does not explore QAML implications for reinforcement learning-based navigation, which forms the core of many autonomous vehicle control stacks.

Our proposed framework advances the state-of-the-art by introducing the first integrated quantum AI pipeline for autonomous vehicles that combines QNN sensor fusion with Nav-Q reinforcement learning under post-quantum security. This integration addresses the limitations in existing work by providing: (1) quantum neural network architectures specifically designed for multi-modal sensor fusion rather than classical preprocessing, (2) Nav-Q quantum reinforcement learning that operates on quantum-fused sensor states rather than classical inputs, and (3) comprehensive post-quantum security that protects the entire QAI pipeline rather than isolated communication channels.

\subsection{Post-Quantum Cryptography in Automotive Systems}

Post-quantum cryptography has gained traction as PQC schemes resist threats posed by large-scale quantum computers. In automotive contexts, Twardokus et al. \cite{haneche2020deep} proposed one of the first practical PQC applications in vehicular networks, focusing on V2V communications using a hybrid authentication protocol combining classical and post-quantum digital signatures for secure Cooperative Awareness Messages (CAMs) without violating IEEE 1609.2 standard latency bounds. Using real-time test beds with vehicle simulators and software-defined radios, they demonstrated PQC-based authentication feasibility in external vehicular communication scenarios.

This work targets external V2V and V2X channels while leaving gaps in protecting internal vehicle communication pathways. With the rise of edge-to-quantum and edge-to-cloud autonomous vehicle architectures, sensor-to-processor communication links remain unprotected under current models, making these channels vulnerable to security attacks and physical tampering. Our framework addresses this overlooked threat by embedding PQC-based encryption and authentication directly into intra-vehicle data pipelines, ensuring quantum-resilient authenticity of sensor streams reaching quantum processor units.

\section{Proposed Framework}

\subsection{System Overview}

Our proposed framework is robust in collecting and processing the sensors data and secures quantum-enhanced navigation communication in autonomous vehicles. This system provides thorough integration of multi-modal sensor fusion, variational quantum processing, adversarial training, and post-quantum secure communication for effective policy execution in both local and vehicle-to-everything contexts.

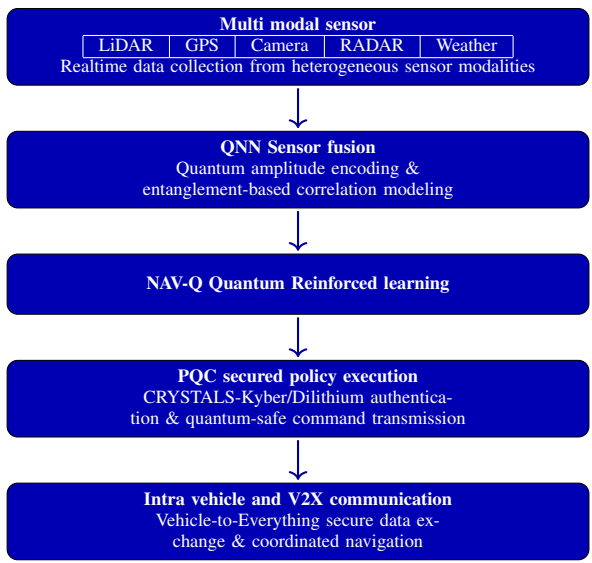
\begin{figure}[!t]
\centering
\begin{tikzpicture}[
    node distance=0.6cm,
    every node/.style={draw, rounded corners, text width=7.5cm, align=center, minimum height=0.8cm, font=\scriptsize},
    sensor/.style={fill=blue!70!black, text=white},
    arrow/.style={->, thick, blue!60!black, shorten >=1pt, shorten <=1pt}
]

% Multi-modal sensor node
\node[sensor] (sensors) {
    \textbf{Multi modal sensor}\\
    \begin{tabular}{|c|c|c|c|c|}
    \hline
    LiDAR & GPS & Camera & RADAR & Weather \\
    \hline
    \end{tabular}\\
    Realtime data collection from heterogeneous sensor modalities
};

% QNN Sensor fusion
\node[sensor, below=of sensors] (qnn) {
    \textbf{QNN Sensor fusion}\\
    Quantum amplitude encoding \&\\
    entanglement-based correlation modeling
};

% NAV-Q
\node[sensor, below=of qnn] (navq) {
    \textbf{NAV-Q Quantum Reinforced learning}
};

% PQC
\node[sensor, below=of navq] (pqc) {
    \textbf{PQC secured policy execution}\\
    CRYSTALS-Kyber/Dilithium authentica-\\
    tion \& quantum-safe command transmission
};

% V2X Communication
\node[sensor, below=of pqc] (v2x) {
    \textbf{Intra vehicle and V2X communication}\\
    Vehicle-to-Everything secure data ex-\\
    change \& coordinated navigation
};

% Arrows
\draw [arrow] (sensors) -- (qnn);
\draw [arrow] (qnn) -- (navq);
\draw [arrow] (navq) -- (pqc);
\draw [arrow] (pqc) -- (v2x);

\end{tikzpicture}
\caption{Quantum AI Pipeline Step1. Architecture Multi-Modal Sensors, Step 2. QNN Sensor Fusion, Step 3. Nav-Q Reinforced Learning, Step4, PQC-Secured Policy Execution, Step5. V2X Communication}
\label{fig:pipeline}
\end{figure}

Our QAI architecture implements a three-stage quantum enabled navigation by proposing quantum advantages at different stages i.e., sensor data collection, data processing, decision making and securing the communication both internal as well as V2X using post quantum cryptographic protocols.

\subsubsection{Multi-Modal Sensor Fusion and Preprocessing}

The system initiates with real-time sensor data collection from multiple sources i.e., LiDAR, radar, camera, GPS, and weather sensors etc., All sensor inputs undergo synchronization and fusion within a preprocessing component which aligns both temporal and spatial frames, for downstream processing. This component handles sensor calibration, noise filtering, and data normalization to prepare inputs for quantum processing.

\subsubsection{Quantum Neural Network (QNN) for Sensor Fusion}

Multi-modal sensor data from the preprocessing stage feeds into a Quantum Neural Network specifically designed for sensor fusion. The QNN employs quantum amplitude encoding to represent high-dimensional sensor correlations in quantum superposition states. This gives exponential expansion of the quantum states compared to classical approaches. This quantum-enhanced fusion process combines LiDAR point clouds, camera images, radar signatures, GPS coordinates, and weather parameters into unified quantum states. This approach captures complex inter sensor relationships which is otherwise impossible to model classically.

\subsubsection{Nav-Q Quantum Reinforcement Learning}

The fused quantum sensor representations from the QNN (3.1.2) feed into a Nav-Q quantum reinforcement learning module. Building on the Nav-Q framework, this component uses variational quantum circuits (VQCs) to learn optimal navigation policies by using quantum policy gradient methods. The quantum Reinforcement Learning agent processes the quantum-fused sensor states and outputs navigation decisions (steering, acceleration, braking) which enables collision-free autonomous driving under complex and dynamic traffic conditions.

\subsubsection{Adversarial Training for Quantum Robustness}

During the training phase, an adversarial training pathway introduces perturbed sensor inputs mimicking realistic attacks (GPS jamming, LiDAR spoofing, camera adversarial patches) to both the QNN sensor fusion and Nav-Q reinforcement learning modules. This helps the integrated QAI system to learn robust policies under adversarial sensor conditions through hybrid quantum-classical adversarial training loops.

\subsubsection{Post-Quantum Cryptographic Security Layer}

All communication within the QAI pipeline right from sensor data transmission to QNN processing, Nav-Q policy computation, and final actuation commands is secured using post-quantum cryptographic protocols. CRYSTALS-Kyber provides quantum-resistant key encapsulation, while CRYSTALS-Dilithium ensures digital signature authenticity. This comprehensive security layer protects against both classical eavesdropping and future quantum cryptanalysis. This approach ensures the integrity of the entire quantum AI navigation system.

\subsubsection{Integrated QAI Policy Execution}

The Nav-Q module outputs navigation policies in quantum state representations that are measured and converted to classical control commands (steering angle, acceleration, braking force). These commands undergo PQC authentication before transmission to vehicle actuators, maintaining sub-50ms latency requirements while ensuring cryptographic integrity throughout the control loop.

\section{Methodology}

\subsection{Quantum Sensor Fusion Module}

\textbf{Amplitude Encoding Scheme:} Using amplitude encoding multi modal sensor data is represented in quantum states. For $n$ sensors producing normalized data vectors $s_i \in \mathbb{R}^{d_i}$, we map each component to quantum amplitudes and construct a composite quantum state:

\begin{equation}
|\psi\rangle = \frac{1}{\sqrt{N}} \sum_{i=1}^{n} \sum_{j=1}^{d_i} \alpha_{i,j} s_{i,j} |i,j\rangle
\end{equation}

where $\alpha_{i,j}$ represents learned attention weights for each sensor component, $s_{i,j}$ denotes the $j$-th normalized component of sensor $i$, and $N$ is a normalization factor ensuring $\langle\psi|\psi\rangle = 1$.

\textbf{Variational Circuit Design:} Using VQC architecture alternating layers of parameterized rotation gates and entanglement operations can be implemented:

\begin{equation}
U(\theta) = \prod_{l=1}^{L} \left[\prod_{q=1}^{Q} R_Y(\theta_{l,q})\right] \prod_{q=1}^{Q-1} \text{CNOT}_{q,q+1}
\end{equation}

where $L$ is circuit depth, $Q$ is qubit count, and $\theta_{l,q}$ are trainable parameters to be optimized using quantum policy gradient methods.

\subsection{Quantum-Adversarial Training Loop}

\textbf{Adversarial Example Generation:} We adapt adversarial training to our hybrid quantum-classical system using projected gradient descent on classical sensor inputs before quantum encoding:

\begin{equation}
s^{(adv)} = \text{Proj}_S(s + \epsilon \cdot \text{sign}(\nabla_s L_{classical}(\theta, s)))
\end{equation}

where $S$ represents the valid sensor input space, $\epsilon$ controls perturbation magnitude, and $L_{classical}(\theta, s)$ denotes the classical loss function computed from quantum measurement outcomes.

\textbf{Robust Loss Function:} Our training objective balances standard policy loss with adversarial robustness:

\begin{equation}
L_{total} = \mathbb{E}_{s\sim D}[L_{policy}(\theta, s)] + \lambda \mathbb{E}_{s^{adv}}[L_{policy}(\theta, s^{adv})]
\end{equation}

where $\lambda$ balances clean and adversarial performance.

\subsection{Post-Quantum Secure Communication Layer}

\textbf{Key Management:} We propose CRYSTALS-Kyber for key encapsulation and CRYSTALS-Dilithium for digital signatures. These algorithms rely on lattice-based mathematical problems which are difficult even for quantum computers and thus addresses the vulnerability of current RSA and elliptic curve cryptography in automotive systems.

\textbf{Communication Protocol:} Our security framework protects all communication within the quantum AI pipeline through a four-step process:

\begin{enumerate}
\item \textbf{Sensor authentication using Dilithium signatures} - Each sensor signs its data to prevent spoofing attacks and ensure authenticity of inputs to the quantum processing pipeline.
\item \textbf{Session key establishment via Kyber key encapsulation} - Quantum-resistant key exchange between sensors and the central QNN processing unit, providing forward secrecy for all communications.
\item \textbf{AES-256 encryption of sensor data using established keys} - All sensor streams (LiDAR, camera, radar, GPS) are encrypted end-to-end using keys derived from the quantum-safe exchange.
\item \textbf{Integrity verification through authenticated encryption} - Message authentication codes detect tampering or corruption, ensuring reliable inputs for quantum reinforcement learning decisions.
\end{enumerate}

This approach secures the entire quantum AI navigation system against both classical and quantum threats while preserving the computational advantages of our proposed framework.

\section{Theoretical Analysis}

\subsection{Advantages of Quantum Neural Network}

\textbf{Exponential Representational Capacity:} Classical neural networks face challenges with the complexity to fuse different sensor types like LiDAR, GPS, RADAR and Camera due to limitation with polynomial representation. QNN solves this problem by effectively correlating states using superposition, thus with $n$ qubits, we can encode $2^n$ states.

\begin{equation}
\text{State Space}_{QNN} = O(2^n) \gg \text{Parameters}_{Classical} = O(\text{poly}(n))
\end{equation}

However, designing quantum circuits and defining measurement processes to get actual information from all possible quantum states, i.e., $2^n$, is a practical challenge.

\textbf{Quantum Sensor Correlation Modeling:} QNN leverages quantum superposition and entanglement within variational circuits to model correlations between different sensor modalities. Through proper parameterization of quantum gates, the system captures complex inter-sensor relationships during training. Quantum superposition states enable cross-modal pattern recognition by encoding correlations between sensor types simultaneously. The parallel processing capabilities of quantum circuits provide advantages for temporal sensor synchronization across multiple data streams. Additionally, quantum coherence properties and measurement processes offer natural noise filtering for more robust feature extraction.

\subsection{Nav-Q Quantum Reinforcement Learning Advantages}

\textbf{Quantum Policy Space Exploration:} Nav-Q's quantum policy representation enables superposition-based exploration of the action space. Theoretical analysis suggests potential for improved convergence in policy optimization, though specific convergence rates depend on problem structure and quantum circuit design \cite{sinha2023nav}.

\textbf{Quantum Advantage in Policy Learning:} The Nav-Q framework provides theoretical guarantees for faster convergence to optimal policies under uncertainty through quantum superposition-based exploration of the policy space. Enhanced exploration emerges from quantum parallel processing of multiple policy candidates simultaneously, while robustness to local minima occurs through quantum interference effects that enable escape from suboptimal policy regions. Variational quantum circuits in reinforcement learning contexts can leverage quantum computational advantages for certain problem structures.

\subsection{Integrated QAI Pipeline Security}

\textbf{Post-Quantum Security Guarantees:} Our PQC implementation provides provable security against quantum adversaries with computational advantages bounded by:

Security Level $\geq 2^{128}$ operations (NIST Level 3)

\textbf{End-to-End Quantum Integrity:} The integrated QNN-Nav-Q-PQC pipeline maintains quantum coherence while ensuring cryptographic security through quantum-safe protocols, providing both computational advantages and security guarantees simultaneously.

\section{Discussion and Limitations}

\subsection{Advantages of Quantum Artificial Intelligence}

\textbf{Integrated QNN-Nav-Q Pipeline:} Our quantum artificial intelligence architecture provides effective integration between sensor fusion and navigation policy learning through quantum state representations which enables end-to-end quantum processing that maintains quantum correlations throughout the pipeline.

\textbf{Potential Exponential Computational Advantages:} In theory, proposed quantum neural networks could encode and process more sensor data patterns than classical methods and the Nav-Q learning system will consistently find better navigation strategies without getting stuck in poor solutions.

\textbf{Quantum-Enhanced Robustness:} Adversarial training techniques proposed in this paper along with probabilistic nature of quantum measurements shall improve robustness of the solution.

\subsection{Limitations}

\textbf{Quantum Hardware Requirements:} Our approach needs quantum processors with enough qubits - probably around 50-100 - and coherence times that can handle real-time automotive demands. Current hardware isn't quite there yet for the millisecond response times that safety-critical driving requires, which is $\sim$50ms latency.

\textbf{QNN Training Complexity:} Training quantum neural networks for multi-modal sensor fusion is challenging in gradient computation and barren plateau avoidance. This will require specialized quantum optimization techniques.

\textbf{Nav-Q Scalability:} While Nav-Q provides theoretical advantages, scaling this to multi-agent complex scenarios in a dynamically changing traffic environments will need progress in the field of quantum multi agent reinforcement learning.

\textbf{Quantum-Classical Interface:} In this paper we discussed about extracting sensor data, processing it and then decision making followed by quantum secure communication. However, in real world, maintaining quantum coherence and interacting with real world all at the same time will have practical challenges to be solved.

\subsection{Future Work}

Our next steps involve finetune the practical implementation steps followed by training the model using simulated or existing data and run on quantum hardware and/or simulators. We plan to adapt our methods for current quantum devices and run experiments using prepared data to validate our approach. We're particularly interested in understanding how quantum advantages translate to measurable improvements by comparing the results with same data in a classical setup.

\section{Conclusion}

This paper presents the pioneering Quantum Artificial Intelligence (QAI) framework tailored for autonomous vehicle navigation, incorporating quantum neural networks for sensor integration, Nav-Q quantum reinforcement learning for policy refinement, and post-quantum cryptography for secure communications. The introduced QAI framework overcomes key constraints in existing autonomous driving systems by leveraging quantum enhancements in perception and decision processes while safeguarding against classical and quantum cyber threats using post-quantum cryptographic methods. The QNN-Nav-Q-PQC workflow establishes a foundational paradigm for secure autonomous navigation, employing quantum artificial intelligence to enhance performance, efficiency, and security. Although the current limitations in quantum hardware preclude immediate real-world application, our research lays down the architectural guidelines and theoretical foundation essential for the future development of quantum-aided transportation systems for autonomous vehicles. The fusion of quantum neural networks with quantum reinforcement learning introduces an innovative perspective in quantum machine learning and autonomous vehicle studies, theoretically illustrating how quantum AI can significantly surpass conventional methods. Future studies will concentrate on short-term quantum applications, hardware validations, and the real-world testing of individual QAI components as quantum technologies advance.

\end{document}